\documentclass[twocolumn,aps,pra]{revtex4}
\usepackage{epsfig}
\usepackage[english]{babel}
\usepackage{latexsym}
\usepackage{graphics}
\usepackage{subfigure}
\usepackage{graphicx}
\usepackage{dcolumn}
\usepackage{amsmath}
\usepackage{hyperref}
\usepackage{amssymb}
\usepackage{color}

\begin{document}
\title{Roles of laser ellipticity in attoclock}

\author{J. Y. Che$^{1,\dag}$, J. Y. Huang$^{1,\dag}$, F. B. Zhang$^{1}$, C. Chen$^{1,*}$, G. G. Xin$^{2}$, and Y. J. Chen$^{1,\ddag}$}

\date{\today}

\begin{abstract}
We study ionization of atoms in strong elliptically-polarized laser fields numerically and analytically.
We focus on effects of laser ellipticity on the offset angle in photoelectron momentum distribution.
This angle is considered to encode time information of tunneling ionization in attoclock experiments.
The calculated offset angle increases with the decrease of ellipticity but the momentum along the major axis of laser polarization
related to this angle  changes slowly, in agreement with experiments.
With a Coulomb-included strong-field model, the scaling laws for ellipticity dependence of this angle
and relevant momentum components are obtained,
and the ellipticity dependence of Coulomb-induced ionization time lag encoded in this angle is also addressed.

\end{abstract}

\affiliation{1.College of Physics and Information Technology, Shaan'xi Normal University, Xi'an, China\\
2.School of Physics, Northwest University, Xi'an, China}
\maketitle

\section{Introduction}
The development of ultrafast and ultrastrong laser technology provides the opportunity for probing the motion of the electron
inside an atom or a molecule at its natural scale \cite{Krausz,Krausz2009,Schultze,Maquet,Vrakking,Pazourek,Vos}.
Relevant probing procedures use photoelectron spectra \cite{Yang1993, Becker2002} or
harmonic spectra \cite{Corkum, Lewenstein1994} generated by the strong interaction
between laser and matter. Attoclock is a such procedure which uses the offset angle in photoelectron momentum distribution (PMD) generated by  ionization of the gas target in strong elliptical laser fields to probe tunneling dynamics \cite{Eckle1,Eckle2,Eckle4}.
Many actual and numerical experiments
on attoclock have been performed for different targets and laser parameters \cite{Landsman,Camus,Klaiber,Torlina,Teeny,Undurti,Quan}.
Some interesting parameter-dependent phenomena have been revealed.
For example, it is showed that the offset angle increases with the decrease of laser intensity \cite{Eckle3}. This typical phenomenon
has attracted great theoretical attention in recent years.
Many efforts have been devoted to the development of applicable Coulomb-included strong-field models to quantitatively explain this phenomenon.
Relevant studies provide deep insights into different aspects of strong-laser-induced tunneling, including
time delay of the tunneling electron under the barrier \cite{Eckle3},
nonadiabatic effects \cite{Boge} and classical scattering \cite{Bray} at the tunnel exit, etc..

Besides of laser intensity,  laser wavelength and ellipticity can also play a nontrivial role in attoclock.
Due to the uncertain in calibrating the laser intensity in experiments,
studies on ellipticity-dependent phenomena in attoclock provide a beneficial complement for intensity-dependent ones \cite{Boge}.
Recent experiments have reported ellipticity-resolved studies on momentum distributions generated from strong-field ionization
of He \cite{Pfeiffer2012,Landsman2013}.
These studies focus on the influence of the Coulomb potential on the longitude momentum spread of the electron wave packet at the tunnel exit
\cite{Pfeiffer2012} and on the momenta related to the offset angle in PMD \cite{Landsman2013}.
It is showed that the studied Coulomb effects are more remarkable for cases of small ellipticity than high ones.
In comparison with cases of laser intensity,
systematic experimental and theoretical studies
on effects of laser ellipticity and wavelength on attoclock are relatively less at present.

In this paper we study ionization of the He atom in strong elliptically-polarized laser fields theoretically.
Through changing laser ellipticity at different laser wavelengths,
we explore the ellipticity-related effects on the offset angle in PMD.
This angle is the main observable which is used for deducing the time-domain information of tunneling in attoclock.
Our simulations are performed through numerical solution of
 time-dependent Schr\"{o}dinger equation (TDSE) and we describe single-electron ionization dynamics
in both two-dimensional (2D) and three-dimensional (3D) cases.

Our results show that when the laser ellipticity increases, the offset angle becomes smaller. For a certain ellipticity,
this angle is larger for shorter laser wavelengths. By contrast, for the momentum components related to the offset angle,
the component along the major axis of laser polarization (denoted by $p_x$ in the paper) is not sensitive to the ellipticity and the wavelength,
suggesting that these angle-related phenomena
arise from the dependence of the momentum component along the minor polarization axis (denoted by $p_y$ here) on ellipticity and wavelength.
These ellipticity-dependent phenomena are well described by a strong-field response-time model
which attributes the complex Coulomb effect into an ionization time lag (i.e., the response time of the electron to light).
With this model, we are able to quantitatively analyze roles of ellipticity in the momentum ($p_x,p_y$) and the offset angle,
as well as in the Coulomb-induced ionization time lag which is mainly related to the momentum component $p_x$  and is encoded in the offset angle.

\section{Theory methods}
\emph{Numerical methods}.-In our simulations, we choose the He atom as the target. In the single-active electron approximation and the length gauge, the Hamiltonian of the model He system interacting with a strong laser field can be written as (in atomic units of $ \hbar = e = m_{e} = 1$)
\begin{equation}
{H}(t)= H_{0}+\mathbf{E}(t)\cdot \mathbf{r}
\end{equation}
Here, $H_{0}={\mathbf{{p}}^2}/{2}+V(\mathbf{r})$ is the field-free Hamiltonian and $V(\textbf{r})=-Z/\sqrt{r^2+\xi}$
is the Coulomb potential with the effective charge $Z$ and the soft-core parameter $\xi$.
The term $\mathbf{E}(t)$ denotes the electric field of the laser. In 2D cases, we have used the
parameters of $Z=1.45$  and $\xi=0.5$.
With these parameters, the ionization potential of the model system reproduced here is $I_p=0.9$ a.u..
In 3D cases, those parameters used are $Z=\sqrt{2I_p}\approx1.34$ and $\xi=0.071$.

In elliptically-polarized cases, the electric field $\mathbf{E}(t)$ used here has the  form of $\mathbf{E}(t)=f(t)[\vec{\mathbf{e}}_{x}E_{x}(t)+\vec{\mathbf{e}}_{y}E_{y}(t)]$, with $E_{x}(t)={E_0}\sin(\omega t)$
and $E_{y}(t)={E_1}\cos(\omega t)$, ${E_0}={E_L}/{\sqrt{1+\varepsilon^2}}$ and ${E_1}=\varepsilon{E_L}/{\sqrt{1+\varepsilon^2}}$.
Here, $E_L$ is the  maximal laser amplitude corresponding to the peak intensity $I$, $\varepsilon$ is the ellipticity,
$\omega$ is the laser frequency and $f(t)$ is the envelope function. The term $\vec{\mathbf{e}}_{x}$($\vec{\mathbf{e}}_{y}$)
is the unit vector along the $x(y)$ axis. We use trapezoidally shaped laser pulses with a total duration of fifteen cycles,
which are linearly turned on and off for three optical cycles, and then kept at a constant intensity for nine
additional cycles. The TDSE of $i\dot{\Psi}(\textbf{r},t)=$H$(t)\Psi(\textbf{r},t)$ is solved numerically
using the spectral method \cite{Feit} with a time step of $\triangle t=0.05$ a.u..
In 2D cases, we have used a grid size of $L_x\times L_y=409.6\times 409.6$ a.u. with space steps
of $\triangle x=\triangle y=0.4$ a.u.. In 3D cases,
the grid size used here is $L_x\times L_y\times L_z=358.4\times 358.4\times 51.2$ a.u.
with $\triangle x=\triangle y=0.7$ a.u. and $\triangle z=0.8$ a.u..
The numerical convergence is checked by using a finer grid.

In order to avoid the reflection of the electron wave packet from the boundary and obtain the momentum space wave function, the coordinate
space is split into the inner and the outer regions with
${\Psi}(\textbf{r},t)={\Psi}_{in}(\textbf{r},t)+{\Psi}_{out}(\textbf{r},t)$, by multiplication using a mask function
$F(\mathbf{r})$. In 2D cases, the mask function has the form of
$F(\mathbf{r})=F(x,y)=\cos^{1/2}[\pi(r_b-r_f)/(L_r-2r_f)]$ for $r_b\geq r_f$ and $F(x,y)=1$  for $r_b< r_f$.
Here, $r_b=\sqrt{x^2+y^2/\epsilon^2}$, $r_f=2.1x_q$ with $x_q=E_0/\omega^2$  and $L_r/2=r_f+50$ a.u. with $L_r\leq L_x$.
The above procedure considers the factors that the quiver amplitude of the ionized electron differs for different laser
parameters and for $x$ and $y$ directions. In 3D cases, the mask function used  is $F(\mathbf{r})=F_1(x,y)F_2(z)$.
The expression of $F_1(x,y)$ is similar to $F(x,y)$ used in 2D cases.
The expression of $F_2(z)$ is $F_2(z)=\cos^{1/2}[\pi(|z|-r_z)/(L_z-2r_z)]$ for $|z|\geq r_z$ and $F_2(z)=1$  for $|z|< r_z$.
Here, $r_z=19.2$ a.u. is the absorbing boundary along the $z$ direction.
In the inner region, the wave function ${\Psi}_{in}(\textbf{r},t)$ is propagated
with the complete Hamiltonian $H(t)$. In the outer region, the time evolution of the wave function ${\Psi}_{out}(\textbf{r},t)$ is carried out
in momentum space with the Hamiltonian of the free electron in the laser field.
The mask function is applied at each time  interval  of 0.5 a.u. and the obtained new fractions of the outer wave function are added to the momentum-space wave function $\tilde{{\Psi}}_{out}(\textbf{r},t)$ from which we obtain the PMD.
Then we find the local maximum in the upper half plane of the PMD and the offset angle $\theta$ is obtained with the local maximum.

\emph{Analytical methods}.-To analytically study the ionization of atoms in strong elliptical laser fields, we use the model termed
as tunneling-response-classical-motion (TRCM) model which gives an applicable description for the intensity-dependent offset angle \cite{Chen2021}. The TRCM arises from strong-field approximation (SFA)
\cite{Lewenstein1995} but considers
the Coulomb effect \cite{MishaY,Goreslavski,yantm2010}.

\emph{SFA description}.-Firstly, according to the SFA with the saddle-point method \cite{Lewenstein1995,Becker2002},
strong-field ionization is characterized by tunneling and each photoelectron drift momentum $\textbf{p}$ has a corresponding
tunneling-out time $t_0$, agreeing with the following mapping relation
\begin{equation}
\mathbf{p}\equiv\mathbf{p}(t_0)=\textbf{v}(t_{0})-\textbf{A}(t_{0}).
\end{equation}
Here, $\textbf{A}(t)$ is the vector potential of the electric field $ \mathbf{E}(t) $.
The tunneling-out time $t_0$ is the real part of the complex time $t_s=t_0+it_x$ that satisfies the saddle-point equation
$[\textbf{p}+\textbf{A}(t_s)]^2/2=-I_p$.
Without considering the Coulomb potential, the tunneling-out time $t_0$ also corresponds to the ionization time at which
the electron is free.  The term $\textbf{v}(t_{0})=\mathbf{p}+\textbf{A}(t_{0})$
denotes the exit velocity of the photoelectron at the exit position (i.e., the tunnel exit)
$\mathbf{r}_0\equiv\mathbf{r}(t_0)=Re(\int^{t_0}_{t_0+it_{x}}[\mathbf{p}+\mathbf{A}(t')]dt')$ \cite{yantm2010}.
This velocity reflects the basic quantum effect of tunneling.
The momentum-time pair ($\textbf{p},t_0$) has been termed as electron trajectory.
The corresponding complex amplitude for the trajectory ($\textbf{p},t_0$)
can be expressed as $c(\textbf{p},t_0)\sim e^{b}$.
Here, $b$ is the imaginary part of the quasiclassical action
$S(\textbf{p},t_s)=\int_{t_s}\{{[\textbf{p}+\textbf{A}(t'})]^2/2+I_p\}dt'$ with $t_s=t_0+it_x$ \cite{Lewenstein1995}.

\emph{TRCM description}.-Around the tunnel exit $r_0=|\mathbf{r}_0|$ which is about 10 a.u. away from the nucleus
for general laser and atomic parameters used in experiments,
the high-energy bound eigenstate of  H$_0$  has large probability amplitudes. The TRCM therefore assumes that
for an actual atom with long-range Coulomb potential, at the tunnel exit $\mathbf{r}(t_0)$, the tunneling electron
with the drift momentum $\textbf{p}$ predicted by SFA  is still located at a quasi-bound state.
This state is characterized by an electron wave packet consisted of high-energy bound eigenstates of  H$_0$,
and approximately agrees with  the virial theorem.
A small period of time $\tau$ is needed for the tunneling electron to evolve
from the quasi-bound state into an ionized state. Then it is free
at the time $t_i=t_0+\tau$ with the Coulomb-included drift momentum $\textbf{p}'$.
This time $\tau$ can be understood as the response time of the electron to light in laser-induced photoelectric effects and is manifested as the Coulomb-induced ionization time lag in strong-field ionization \cite{Xie,Wang2020}.
The mapping between the drift momentum $\textbf{p}'$ and the ionization time $t_i$ in TRCM is expressed as
\begin{equation}
\mathbf{p}'\equiv\mathbf{p}'(t_i)=\textbf{v}(t_{0})-\textbf{A}(t_{i}).
\end{equation}

\emph{Angle formula}.-The offset angle $\theta$ in PMD is related to the most probable route (MPR)
which corresponds to the momentum having the maximal amplitude in PMD. For MPR, the tunneling-out time $t_0$ of  the
photoelectron agrees with the peak time of the laser field. This angle $\theta$ satisfies the following relation
\begin{equation}
\tan\theta=p'_x/p'_y= A_x(t_i)/[A_y(t_i)-v_y(t_0)].
\end{equation}
The above expression has considered the factor that for the MPR $v_x(t_0)=0$.
By neglecting $v_y(t_0)$,  the adiabatic version of the above expression is also obtained. That is
\begin{equation}
\tan\theta\approx A_x(t_i)/A_y(t_i).
\end{equation}
The adiabatic version is applicable for $\gamma\ll1$ with  $\gamma=w\sqrt{2I_p}/E_0$
being the Keldysh parameter \cite{Keldysh}. It has been used in  \cite{Che2} to deduce the lag $\tau$
of the asymmetric system HeH$^{+}$ and has been termed as Coulomb-calibrated attoclock (CCAC).
In CCAC, with considering  $t_{i}=t_{0}+\tau $ and $\omega t_{0}=\pi/2$ for MPR,
we can further obtain the following relation
\begin{equation}
\tan\theta\approx\tan(\omega\tau)/\varepsilon\approx\theta\approx\omega\tau/\varepsilon
\end{equation}
for a small angle $\theta$. With these above expressions, when the lag $\tau=t_{i}-t_{0}$ is obtained analytically or numerically,
one can further obtain the offset angle $\theta$. In turn,
when the  angle $ \theta $ is obtained in experiments or TDSE simulations, one can also deduce the lag $\tau$ from this angle.

\emph{Lag formula}.-According to the assumptions in TRCM, at the tunneling-out time $t_0$, the electron is still located
in a quasi-bound state $\psi_b$ which approximately agrees with the virial theorem.
The average potential energy of this state is
$\langle V(\mathbf{r})\rangle\approx V(\textbf{r}(t_0))$ and the average kinetic energy is
$\langle\textbf{v}^2/2\rangle=n_f\langle v_x^2/2\rangle\approx-V(\textbf{r}(t_0))/2$.
This Coulomb-induced velocity $|v_x|\approx\sqrt{|V(\mathbf{r}(t_0))|/n_f}$
reflects the basic symmetry requirement of the Coulomb potential on the electric state.
A time lag $\tau$ is needed for the tunneling electron to acquire an impulse from the laser field in order to break this symmetry.
Then for the MPR, the lag $\tau$ can be evaluated with the expression of $\tau\approx\sqrt{|V(\mathbf{r}(t_0))|/n_f}/E_0$.
Here, $n_{f}=2,3$ is the dimension of the system studied and the exit position
$\mathbf{r}(t_0)$ is determined by the saddle points as discussed above.
For a hydrogen-like atom with the form of the Coulomb potential $V(\textbf{r})=-Z/r$,
by neglecting the field $E_y(t)$ in the solution of the saddle-point equation,
an approximate expression for $\tau$ can also be obtained. That is
\begin{equation}
\tau\approx\sqrt{Z\omega^2/[n_fE_0^3(\sqrt{\gamma^2+1}-1)]}.
\end{equation}
For real 3D cases such as in experiments, the value of the effective charge $Z$ can be evaluated with $Z=\sqrt{2I_p}$.
For 2D TDSE, the value of $Z$ can be chosen as that used in simulations.
Similarly, an approximate expression for $v_y(t_0)$ can also be obtained. That is
\begin{equation}
v_y(t_0)=[\varepsilon\sqrt{2I_p}/\text{arcsinh}(\gamma)-E_1/\omega]\sin\omega t_0.
\end{equation}
By inserting Eq. (7) and Eq. (8) into Eq. (4), we can analytically evaluate the offset angle $\theta$.
In this paper, we denotes the above manner for obtaining $\theta$ TRCM.
Similarly, by inserting  Eq. (7) into Eq. (6), the adiabatic prediction of the angle $\theta$ can also be obtained,
and this manner can be denoted as CCAC.

\emph{PMDs}.-In TRCM treatment, by assuming that for an arbitrary SFA electron trajectory ($\textbf{p},t_0$),
the Coulomb potential
 does not influence the corresponding complex amplitude  $c(\textbf{p},t_0)$,
we can obtain the TRCM amplitude  $c(\textbf{p}',t_i)$ for Coulomb-included electron trajectory ($\textbf{p}',t_i$) directly from the SFA one
with $c(\textbf{p}',t_i)\equiv c(\textbf{p},t_0)$ at $\tau\approx\sqrt{|V(\mathbf{r}(t_0))|/n_f}/|\textbf{E}(t_0)|$.
This TRCM therefore allows the analytical evaluation of the Coulomb-included PMD without the need of solving the Newton equation including
both the electric force and the Coulomb force.
The TRCM prediction of the PMD for He is presented in the right column of Fig. 1.

%%%%%%%%%%%%%%%%%%%%%%%%%%%%%%%%%%%%%%%%%%%%%%%%%%%%%%%%%%%%%%%%%%%%%%%%%%%%%%%%%%%%%%%%%%%%%
%%%%%%%%%%%%%%%%%%%%%%%%%%%%%%%%%%%%%%%%%%%%%%%%%%%%%%%%%%%%%%%%%%%%%%%%%%%%%%%%%%%%%%%%
\begin{figure}[t]
\begin{center}
\rotatebox{0}{\resizebox *{6.5cm}{9cm} {\includegraphics {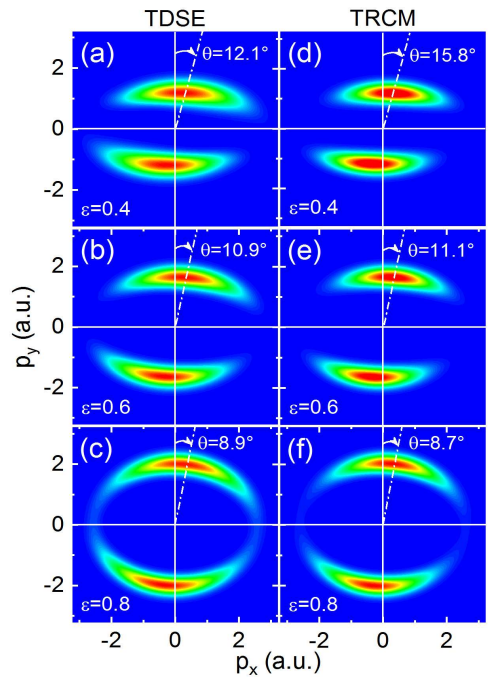}}}
\end{center}
\caption{PMDs of He obtained with 2D TDSE (left column) and TRCM (right) at different laser ellipticity $\varepsilon$.
Laser parameters used are $I=7\times10^{14}$ W/cm$^{2}$ and $\lambda=1000$ nm. The laser ellipticity is as shown.
The offset angle $\theta$ relating to the momentum with the maximal amplitude in PMD  is also indicated in each panel.
}
\label{fig:g1}
\end{figure}
%%%%%%%%%%%%%%%%%%%%%%%%%%%%%%%%%%%%%%%%%%%%%%%%%%%%%%%%%%%%%%%%%%%%%%%%%%%%%%%%%%%%%%%%%%
%%%%%%%%%%%%%%%%%%%%%%%%%%%%%%%%%%%%%%%%%%%%%%%%%%%%%%%%%%%%%%%%%%%%%%%%%%%%%%%%%%%%%%%%%
\begin{figure}[t]
\begin{center}
\rotatebox{0}{\resizebox *{8cm}{8.5cm} {\includegraphics {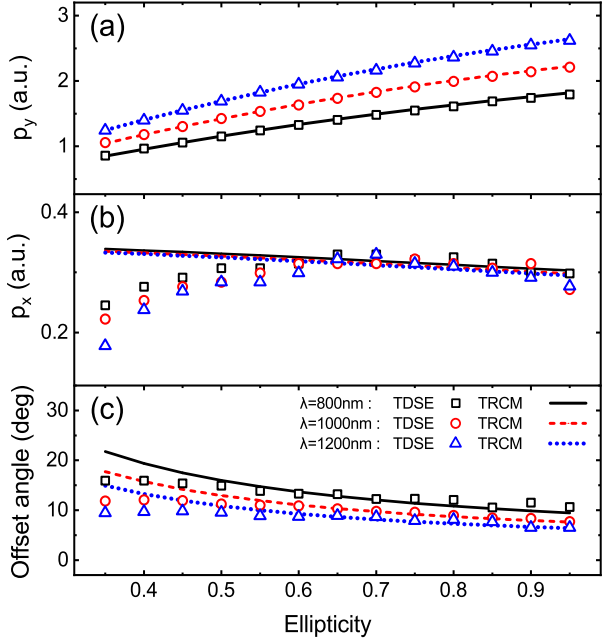}}}
\end{center}
\caption{Comparisons of the drift momentum ($p_x,p_y$) related to MPR and the offset angle $\theta$ between 2D TDSE and TRCM of Eq. (4)
at different laser ellipticity $\varepsilon$ and wavelengths $\lambda$.
The laser intensity used is  $I=7\times10^{14}$ W/cm$^{2}$.
}
\label{fig:g2}
\end{figure}
%%%%%%%%%%%%%%%%%%%%%%%%%%%%%%%%%%%%%%%%%%%%%%%%%%%%%%%%%%%%%%%%%%%%%%%%%%%%%%%%%%%%%%%%%%

\section{Results and discussions}
\emph{2D cases}.-We first present our comparisons between TDSE and TRCM in 2D cases which allow us to explore a wide parameter region.
In Fig. 1, we present PMDs of He obtained by TDSE and TRCM in strong elliptical laser fields with different $\varepsilon$.
The offset angle in PMD of TDSE decreases with the increase of ellipticity, as seen in the left column of Fig. 1.
This decreasing trend is well reproduced by TRCM, as seen in the right column of Fig. 1.
For relatively large ellipticity of $\varepsilon\geq0.6$,
the TRCM offset angle in PMD is also in quantitative agreement with the corresponding TDSE one.
For the case of $\varepsilon=0.4$ in the first row of Fig. 1, the TRCM result is about 4 degrees larger than the TDSE one,
suggesting that the TRCM is more applicable for the case of relatively large ellipticity. We will return to this point later.

\emph{Roles of $\varepsilon$}.-In Fig. 2, we further show the comparisons for the offset angle
and the momentum ($p_x,p_y$), which is associated with this angle
and has the maximal amplitude in PMD, at diverse
laser ellipticity and wavelengths.
Without loss of generality, in this paper, we consider the cases of the momentum ($p_x,p_y$)
being located in the first quadrant of PMD with $p_x>0, p_y>0$.

Firstly, for the momentum $p_y$ along the minor axis of polarization,
one can see from Fig. 2(a), the TDSE and TRCM results agree well with each other.
The TDSE and TRCM predictions of $p_x$ along the major axis of polarization deviate remarkably from each other
for the case of small ellipticity of $\varepsilon\leq0.4$ and
become near to and coincident with each other for intermediate and larger ellipticity, as seen in Fig. 2(b).
Accordingly, in Fig. 2(c), the angles predicted by TDSE and TRCM also show the remarkable difference for small ellipticity and
agree with each other at larger ellipticity with $\varepsilon\ge0.5$.
This remarkable difference  between TRCM and TDSE can arise from the following reason.
For small ellipticity, the ionized electron wave packet related to a rescattering electron trajectory can interfere
with the wave packet related to a direct-ionization electron trajectory \cite{Becker2002}.
This interference will influence the amplitude of PMD and
therefore influence the identification of the offset angle.
These influences are not considered in TRCM.

It is also worth noting that in Fig. 2, for the region of $0.6\leq\varepsilon\leq0.95$ at which the TRCM works well,
the momentum $p_x$ decreases slowly with the increase of ellipticity.
By contrast, the momentum $p_y$ increases remarkably with increasing the ellipticity.
One therefore can expect that in this region, the dependence of the momentum $p_y$ on ellipticity  plays a more important role in
the ellipticity dependence of the offset angle $\theta\approx p_x/p_y$.

\emph{Scaling laws}.-Through TRCM, we can also obtain the different scaling laws
for the dependence of these two momentum components $p_x$ and $p_y$ on the
ellipticity $\varepsilon$. By Eq. (5) of the adiabatic approximation which is applicable for a small Keldysh parameter, we have $p_y\approx A_y(t_0)\approx \varepsilon E_0/\omega$ and $p_x\approx A_x(t_i)\approx E_0\tau\approx\sqrt{|V(\mathbf{r}(t_0))|/n_f}$. By considering that $V(r)=-Z/r$, $r(t_0)\approx I_p/E_0$, and
$E_0=E_L/\sqrt{1+\varepsilon^2}$, we also have
\begin{equation}
p_x\approx\sqrt{ZE_L/(n_fI_p\sqrt{1+\varepsilon^2})}= A(1+\varepsilon^2)^{-1/4}
\end{equation}
and
\begin{equation}
p_y\approx \varepsilon E_L/(\omega\sqrt{1+\varepsilon^2})= B\varepsilon(1+\varepsilon^2)^{-1/2}.
\end{equation}
Here, $A\equiv A(E_L,I_p)=\sqrt{ZE_L/(n_f{I_p})}$ with $Z=\sqrt{2I_p}$ which is independent of the frequency $\omega$  and $B\equiv B(E_L,\omega)=E_L/\omega$ which is independent of the ionization potential $I_p$.
Then we obtain the scaling laws for the momentum ($p_x,p_y$) with
\begin{equation}
p_x\sim (1+\varepsilon^2)^{-1/4}; p_y\sim \varepsilon(1+\varepsilon^2)^{-1/2}.
\end{equation}
For the parameter region of $0.3\leq\varepsilon\leq1$ explored in this paper, we have $0.98A\geq p_x\geq0.84A$ and $0.29B\leq p_y\leq0.71B$,
which indicate the slow decrease of  $p_x$ and the remarkable increase of $p_y$ with the increase of $\varepsilon$ observed in Fig. 2(b) and
Fig. 2(a), respectively. Equations (9) and (10) also shed light on different responses of $p_x$ and $p_y$
to the laser wavelength $\lambda$ seen in Fig. 2.

By Eq. (9) and Eq. (10), we can also obtain the scaling law for the ellipticity-dependent angle
$\theta$ with $\theta\approx p_x/p_y$ and $p_y\neq0$. That is
\begin{equation}
\theta\approx C(1+\varepsilon^2)^{1/4}\varepsilon^{-1}\sim(1+\varepsilon^2)^{1/4}\varepsilon^{-1}.
\end{equation}
Here, $C\equiv C(E_L,I_p,\omega)=\omega\sqrt{Z/(n_f{I_p}E_L)}$. Equation (12) provides explanations for ellipticity-wavelength-related
phenomena in Fig. 2(c).

Considering $p_x\approx  E_0\tau$ and Eq. (9), we can also obtain the relevant scaling law for the lag $\tau$. That is
\begin{equation}
\tau\approx D(1+\varepsilon^2)^{1/4}\sim(1+\varepsilon^2)^{1/4}.
\end{equation}
Here,  $D\equiv D(E_L,I_p)=\sqrt{Z/(n_f{I_p}E_L)}$. Equation (13) shows that the lag slowly increases with increasing the ellipticity and
can be used to analyze ellipticity-dependent phenomena for the lag $\tau$, as to be shown in Fig. 3.

\emph{Time lag}.-In Fig. 3, we further compare the ionization time lag obtained with different methods.
Not as the offset angle which is related to the PMD and therefore
can be directly measured in experiments,
the lag is related to the instantaneous ionization property of the laser-driven system
which is not easy to probe in experiments.  However, in TDSE simulations,
this lag can be evaluated with approximately calculating the instantaneous ionization rate.
Specifically, we first find the time $t_{i}$ which corresponds to the maximal value of the instantaneous ionization rate $P(t)=dI(t)/dt$ \cite{Xie}. Here, $I(t)=1-\sum_{n} |\langle n|\Psi(\textbf{r},t)\rangle |^{2} $ is the instantaneous ionization yield, $|n\rangle$ is the bound eigenstate of
the field-free Hamiltonian $H_{0}$ and $|\Psi(\textbf{r},t)\rangle$ is the TDSE wave function of $H(t)$.
We only consider the first several bound eigenstates with $n$=0,1,2...5. The upper limit $n_{u}$ of $n$ is  determined with the eigenenergy $E_{n_{u}+1} $ of the $(n_{u}+1)$th eigenstate approximately agreeing with the semiclassical analysis. That is
$E_{n_{u}+1}\approx V(\textbf{r}(t_{0}))+v_x^{2} /2 $ with $|v_x|=\sqrt{|V(\mathbf{r}(t_0))|/n_f}$.
Then the lag $\tau$ is obtained with $\tau=t_{i}-t_{0}$
where $t_0$ is the neighboring peak time of the laser field agreeing with $|E_{x}(t_{0})|=E_{0}$.
We mention that the value of $t_i$ evaluated here depends on
the number $n'=n_u+1$ of the bound states excluded from $|\Psi(\textbf{r},t)\rangle$. Our simulations show that when the number $n'$ is larger,
the time lag obtained is also larger. However, despite this dependence, the appearance of a nonzero $\tau$ at different numbers $n'$
suggests that at a certain time $t_0$,
the electronic wave packet which leaves the ground state  does not appear at the continuum state instantly and therefore
the maximal ionization rate also does not appear at the peak time of the laser field. In particular, as to be shown in the following,
with the similar definition of ionization to TRCM, the TDSE prediction of $\tau$ is comparable to the TRCM one.

In addition to comparing the TDSE results to the TRCM predictions of Eq. (7), we also compare them to the CCAC predictions of Eq. (6) with  $\tau\approx\varepsilon\theta/\omega$, where the offset angle $\theta$ is obtained from PMD of TDSE.

%%%%%%%%%%%%%%%%%%%%%%%%%%%%%%%%%%%%%%%%%%%%%%%%%%%%%%%%%%%%%%%%%%%%%%%%%%%%%%%%%%%%%%%%%
\begin{figure}[t]
\begin{center}
\rotatebox{0}{\resizebox *{8cm}{5cm} {\includegraphics {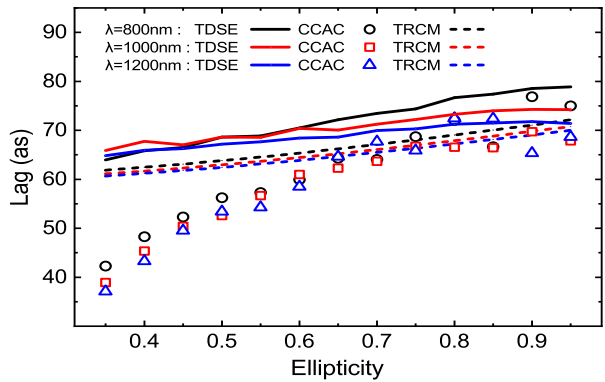}}}
\end{center}
\caption{Comparisons of the time lag $\tau$ between 2D TDSE, CCAC
and TRCM of Eq. (7) at different laser ellipticity $\varepsilon$ and wavelengths $\lambda$.
The TDSE results are obtained by finding the peak time of the instantaneous ionization rate.
The CCAC results are obtained by deducing the lag $\tau$ from the TDSE offset angle $\theta$ with
$\tau\approx\varepsilon\theta/\omega$ of Eq. (6).
The laser intensity used is $I=7\times10^{14}$ W/cm$^{2}$.
}
\label{fig:g3}
\end{figure}
%%%%%%%%%%%%%%%%%%%%%%%%%%%%%%%%%%%%%%%%%%%%%%%%%%%%%%%%%%%%%%%%%%%%%%%%%%%%%%%%%%%%%%%%%

Firstly, one can observe from Fig. 3 that the TDSE results of the lag $\tau$ increase with the increase of ellipticity. In addition,
the TDSE results for different wavelengths at a certain value of ellipticity are comparable when $\varepsilon\leq0.6$,
and they begin to differ somewhat from each other for $\varepsilon>0.6$ with the case of shorter wavelengths showing larger $\tau$.
The trends for  both the parameter regions of  $\varepsilon\leq0.6$ and $\varepsilon>0.6$ are well reproduced by the TRCM.
In particular, the quantity of the TDSE lag is comparable to the TRCM one with a difference smaller than 10 attoseconds.
These ellipticity-dependent phenomena for the angle $\theta$ can also be understood with Eq. (13). It should be noted that Eq. (13) is independent
of the laser wavelength $\lambda$, but these TDSE and TRCM results in Fig. 3 show a weak dependence on $\lambda$ for larger $\varepsilon$.
This reason is that Eq. (13) uses the approximation of $r(t_0)\approx I_p/E_0$ which neglects the effect of laser wavelength.
By comparison,  Eq. (7), which is used to obtain the TRCM results shown here, includes the influence of $\lambda\sim1/\omega$.

%%%%%%%%%%%%%%%%%%%%%%%%%%%%%%%%%%%%%%%%%%%%%%%%%%%%%%%%%%%%%%%%%%%%%%%%%%%%%%%%%%%%%%%%%
\begin{figure}[t]
\begin{center}
\rotatebox{0}{\resizebox *{8cm}{8.5cm} {\includegraphics {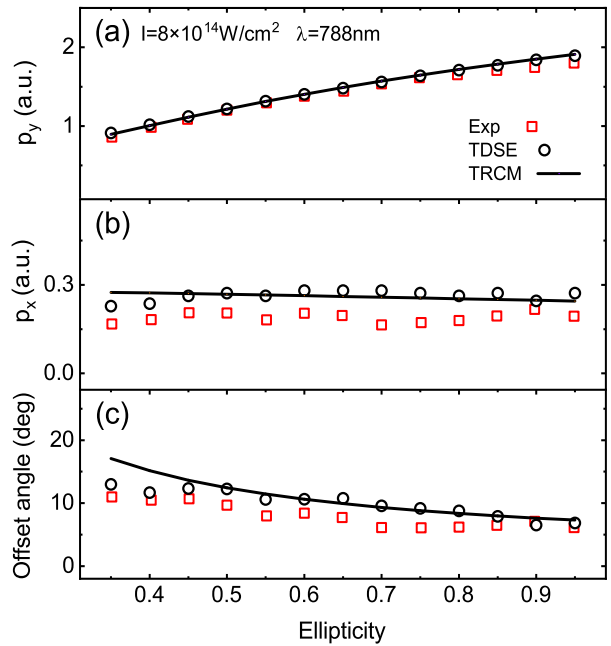}}}
\end{center}
\caption{Comparisons of the drift momentum ($p_x,p_y$) related to MPR and the offset angle $\theta$
between experiments  \cite{Landsman2013}, 3D TDSE and TRCM of Eq. (4)
at different laser ellipticity $\varepsilon$.
In (c), the experimental offset angles  are obtained with $ \theta\approx\arctan(p_{x} /p_{y})$.
Laser parameters used are $I=8\times10^{14}$ W/cm$^{2}$ and $\lambda=788$ nm.
}
\label{fig:g4}
\end{figure}
%%%%%%%%%%%%%%%%%%%%%%%%%%%%%%%%%%%%%%%%%%%%%%%%%%%%%%%%%%%%%%%%%%%%%%%%%%%%%%%%%%%%%%%%%
%%%%%%%%%%%%%%%%%%%%%%%%%%%%%%%%%%%%%%%%%%%%%%%%%%%%%%%%%%%%%%%%%%%%%%%%%%%%%%%%%%%%%%%

Secondly, the CCAC results differ from the TDSE and TRCM ones for $\varepsilon\leq0.6$ and
become near to them for $\varepsilon>0.6$. Since the CCAC lag $\tau$ is obtained with the TDSE offset angle,
these angle differences between TDSE and TRCM in Fig. 2(c) for smaller ellipticity also shed light on
the corresponding lag differences between CCAC and TRCM in Fig. 3.

Thirdly, not as the remarkable difference between TDSE and TRCM for the momentum component $p_x$ at smaller $\varepsilon$ seen in Fig. 2(b),
the predictions of TDSE and TRCM for the lag are comparable at different ellipticity in Fig. 3.
This suggests that the TRCM is capable of providing a good description for time-resolved ionization dynamics at different laser ellipticity.
The offset angle as well as the momentum $p_x$ related to this angle are possibly influenced by other effects such as quantum interference
between different electron trajectories at small ellipticity \cite{Becker2002}.
As a result, the comparison between TDSE and TRCM for this angle or the momentum $p_x$ can also be influenced.

\emph{3D cases}.-In Fig. 4, we further compare the predictions of TRCM to results of 3D TDSE and experiments.
It is worth noting that by Eq. (7), this lag is smaller in 3D cases than in 2D ones with similar laser and atomic parameters,
so does the offset angle. From Fig. 4(a), one can observe that the results of TRCM, 3D-TDSE and experiments for $p_y$ agree well
with each other. They show that the value of $p_y$ increases remarkably with the increase of ellipticity.
For the case of $p_x$ in Fig. 4(b), the results of TRCM and 3D-TDSE are also in good agreement
with each other for $\varepsilon\geq0.45$, but differ somewhat from each other for  $\varepsilon<0.45$. They also differ from
the experimental results for different ellipticity with a difference of $\triangle p_x\approx0.07$ a.u..
Despite this difference, all of these results in Fig. 4(b) show that the value of $p_x$
almost does not change with the increase of ellipticity, similar to 2D cases.
The comparisons in Fig. 4(c) for the offset angle are somewhat similar to the cases of $p_x$ in Fig. 4(b),
but the angle curves show a clear decreasing trend with the increase of ellipticity in Fig. 4(c).
When the TRCM and 3D-TDSE results are also consistent with each other for $\varepsilon\geq0.45$, they are somewhat higher than
the experimental ones with a difference of about 1 degree to 2 degrees.
The above ellipticity-dependent phenomena  can also be well understood with the scaling laws of Eqs. (11) and (12). 
The difference between TDSE and model results and experiments may partly arise from the uncertain of laser intensity used in experiments. 
Experiments for more targets are highly desired to further validate these ellipticity-dependent phenomena.

\section{Conclusion}
In summary, we have studied the ionization of He in strong elliptical laser fields with different laser
ellipticity and wavelengths numerically and analytically. We have compared the TDSE results to predictions of
a Coulomb-included strong-field model termed as TRCM and to experiments.
The calculated offset angle in photoelectron momentum distribution decreases with
the increase of laser ellipticity, in agreement with the experimental measurement and the model prediction.
This phenomenon can be understood with analyzing the ellipticity dependence of the momentum components $p_x$ and $p_y$ related to this angle.
When this component $p_x$ along the major axis of laser polarization is not sensitive to ellipticity, this component $p_y$ along the minor one
increases remarkably with increasing the ellipticity.
With the TRCM model, the scaling laws of the momentum ($p_x$, $p_y$) and the angle $\theta$ to the ellipticity $\varepsilon$ are also given,
which explain the ellipticity-dependent phenomena for these two momentum components and this angle.
Because the momentum component $p_x$ is closely associated with the Coulomb-induced ionization time lag,
we also further discuss the dependence
of this lag on ellipticity. We evaluate this lag with different methods including TDSE one, TRCM one and a mix of TDSE and TRCM termed as CCAC.
These different methods give similar ellipticity-dependent results for the lag $\tau$ at larger ellipticity, indicating that
one can deduce this lag from the offset angle measured in attoclock experiments.

%\section*{Acknowledgement}

This work was supported by the National Natural Science Foundation of China (Grant No. 12174239),
and the Fundamental Research Funds for the Central Universities of China (Grant No. 2021TS089).

\end{document}